\documentclass[12pt,english]{article}
\usepackage[T1]{fontenc}
\usepackage[latin9]{inputenc}
\usepackage{setspace}
\makeatletter
\usepackage{aStyle}

\makeatother

\usepackage{babel}
\begin{document}
\begin{doublespace}
\begin{center}
\textbf{\Large{}Virgin and Under-Investigated Areas of Research in
Non-Newtonian Fluid Mechanics}\vspace{-1.3cm}
\par\end{center}
\end{doublespace}

\begin{center}
Taha Sochi\footnote{University College London - Department of Physics \& Astronomy - Gower
Street - London - WC1E 6BT. Email: t.sochi@ucl.ac.uk.}\vspace{-0.4cm}
\par\end{center}

\noindent \phantomsection \addcontentsline{toc}{section}{Abstract}

\noindent \textbf{Abstract}: This paper is mainly about the areas
of non-Newtonian fluid mechanics which are not investigated or not
appropriately and sufficiently investigated. In fact, this should
also include emerging areas of research in the field of non-Newtonian
fluid mechanics due to new scientific and technological developments
and advancements. The purpose of the paper is to highlight and draw
the attention to these areas so that researchers (especially the young
researchers and new-comers to research such as PhD students) invest
their resources and efforts in these areas instead of investing in
other areas which are previously investigated and hence they are of
less priority from this aspect. Apart from the obvious benefit of
``leveling up'' in research, the attention to these rather neglected
and non-explored areas of research can be beneficial at the scientific
and individual levels since it can lead to breakthroughs and new discoveries
in these areas of research by inspecting and assessing their potentials
and impacts at the theoretical and practical levels and probing their
beneficial applications. We will also provide a brief discussion about
the possibility of introducing novel tools and methods in these areas
of research (and in non-Newtonian fluid mechanics research in general)
as well as highlighting some of the existing limitations of the past
and current research in the field of non-Newtonian fluid mechanics
(noting that this discussion should help in achieving the ultimate
objective of this investigation).\vspace{0.3cm}


\clearpage{}

\phantomsection \addcontentsline{toc}{section}{Table of Contents}

\tableofcontents{}

\clearpage{}

\section{Introduction}

Non-Newtonian fluid mechanics (and related subjects in mechanics and
rheology) is a huge branch of science. It has been under intensive
investigation and research from many aspects and perspectives especially
in the last decades (starting approximately from the 1950s although
the interest in non-Newtonian flow phenomena goes back well beyond
this date as can be inferred for instance from Maxwell viscoelastic
fluid model). The subject has been investigated from physical, mathematical
and computational angles and by different methodologies. It was also
a subject to many experimental as well as theoretical investigations
(see for instance \cite{BirdbookAH1987,CarreaubookKC1997}). This
should come as no surprise because non-Newtonian flow is a common
physical phenomena in nature (and actually it is the norm rather than
the exception) and has many technological and industrial applications
(e.g. in oil extraction, medicine, engineering, food processing, and
so on).

However, thorough inspection to the previous and current research
in this field should reveal the fact that there are many gaps and
neglected areas in these investigations where some areas are ignored
or insufficiently investigated while other areas enjoyed much more
attention and investments in resources and efforts. This is partly
understandable because (for instance) some areas are easier to investigate
or have exceptional practical benefits (e.g. to the petroleum industry)
and hence they were the focus of many investigations while other areas
are more difficult to investigate or have less practical benefits
and hence they were ignored or under-investigated. We should also
consider in this context newly emerging areas of research in this
field due to new developments in science, medicine and technology.

Anyway, this fact should be admitted and hence addressed by the research
community to level up the situation by directing the attention to
the ignored and under-investigated areas and drawing more resources
and efforts to these areas. This is not only a necessity to science
to have a better and more thorough understanding of the non-Newtonian
flow phenomena but it should also have many scientific and personal
benefits where new areas of research could lead to breakthroughs and
novelties that to be celebrated by science and scientists.

As indicated already in the abstract, the main purpose of this paper
is to draw the attention to these rather neglected or insufficiently
investigated areas of non-Newtonian fluid mechanics to help identifying
these areas first, and to encourage the researchers in this field
to spend more resources and efforts in these areas by directing some
or all of their resources and efforts to these areas instead of accumulating
their attention and resources on the traditional areas of this subject.
In fact, we also want this paper to be a call and motive to the funding
bodies and academic and research institutes (especially the leading
ones) to give these areas more attention in their funding and projects
(such as PhD scholarships and programs and post-doctoral research
projects).

As indicated already, there are many serious challenges in the research
of these virgin and under-investigated areas and hence this could
be a deterrent (or at least makes them less attractive) to many researchers
to enter these areas (especially those who come to research temporarily
and accidentally and for other objectives such as getting formal academic
qualifications). However, this can be a motive to other types of researchers
(and actually the talented and exceptional researchers) who enjoy
the challenges and difficulties and are keen to find areas of research
that can have lasting effects and impacts at the scientific and personal
levels by making breakthroughs and new discoveries.

In short, although there are many challenges, deterrents and disadvantages
in conducting research in these virgin and under-investigated areas,
there are also many benefits, incentives and advantages in conducting
research in these areas (such as dealing with and managing less complexities
in many cases where investigations of novel subjects require little
more than addressing and dealing with fundamentals and generalities,
possibility of making excited discoveries and breakthrough which are
less likely in old areas of research and hence getting precedence
and honors, and so on).

Our plan in this paper (following this introduction) is to present
and discuss some research areas that are not explored or they require
further attention and investigation (see $\S$ \ref{secVirginUnderInvestigated}).
We then (for the purpose of completeness) discuss briefly some of
the potential tools that should be considered in these investigations
(and actually in the investigations of non-Newtonian fluid mechanics
in general) especially with the current revolution in computing and
artificial intelligence (see $\S$ \ref{secNovelTools}). We then
(also for the purpose of completeness) discuss briefly some of the
limitations of the past and current research in non-Newtonian fluid
mechanics (see $\S$ \ref{secLimitationsOfPast}). The paper will
be concluded by listing a number of achievements and conclusions that
we obtain from the present paper.

\section{\label{secVirginUnderInvestigated}Virgin and Under-Investigated
Areas of Research}

We present in the following subsections some of the neglected or emerging
or under-explored topics that should benefit from deeper and more
thorough investigation and hence they should be given more attention
in the future research in the field of non-Newtonian fluid mechanics.
However, before we go through this investigation it is important to
note that the following items of discussion (as outlined by the titles
of the following subsections) are not supposed to be mutually exclusive
but they rather represent various views from various angles to the
subject of non-Newtonian fluid mechanics and hence there are considerable
overlaps (although this should not be considered as repetition due
to the differences in perspectives and considerations). We should
also note that although we try to be as comprehensive as possible
in this discussion (within the scope and size of the present paper),
our investigation is far from being thorough and hence the following
items of discussion should be considered as a good and large sample
of the issues and items in this regard.

\subsection{\label{subsecGeometry}Non-Newtonian Flow in Geometries of Complex
Morphology}

Most of the past and current research in non-Newtonian fluid mechanics
focuses on idealized flows in simple geometries, such as the flow
of simple fluids (like power law fluids) in circular pipes (or thin
slits or channels of simple geometry) under simple and idealized conditions.
However, many real-world applications involve geometries of complex
morphology (such as complex porous media as exemplified by geological
structures, irregular industrial pipes, medical devices, biological
tissues, and so on). As indicated, this issue is not limited to the
geometry of the flow containers and conduits but to the geometry plus
the attached conditions (regarding for example the flow, fluid and
environment) as well as the fluid models which are supposed to represent
the rheology of the fluid which flows in these geometries. For example,
when we talk in the following about the flow in conduits of regular
non-circular cross sections we should consider flow of different non-Newtonian
fluid models (e.g. power law, Ellis, Herschel-Bulkley, Carreau, Cross,
etc.) with different conditions (e.g. with and without slip, compressible
and incompressible, with and without external body forces and fields,
with different boundary conditions, and so on).\footnote{In fact, these considerations (as well as their combinations) are
what makes these areas of research virgin or under-investigated. Otherwise,
most of the areas that we will discuss in the subsections of this
section are investigated in general (although with limitations in
these considerations) and hence we can find considerable amounts of
literature in most of these areas.}

A sample of suggested areas of research in this field is:
\begin{enumerate}
\item Flow in conduits and pipes of regular non-circular cross sections
such as elliptical, rectangular, triangular, circular sector, half
moon (or crescent) and half round, limacon, polygon (convex or concave,
and simple or complex), polygon with curved sides, star polygon, and
so on.
\item Flow in conduits and pipes of irregular cross sections such as some
of the previous geometries (e.g. irregular polygons), totally irregular
geometry (with and without variation in the cross section along the
conduit length), and so on.
\item Flow in conduits of multiple (or non-simple) connectivity such as
circular concentric annulus, circular eccentric annulus, circular
annulus with non-circular hole or holes (e.g. elliptical or square
or triangular hole), some of the previous geometries with concentric/eccentric
circular/non-circular holes (e.g. rectangular with concentric/eccentric
hole or holes of various regular/irregular geometries), and so on.
\item Flow in converging-diverging (or diverging-converging) and corrugated
regular geometries. Some irregularities in these geometries can also
be considered (such as variations in the convergence-divergence and
corrugation parameters). In fact, convergence-divergence and corrugation
should be considered with conduits of various general geometries and
hence it should include circular and non-circular conduits of various
geometric shapes as given in the previous points.
\item Flow in conduits and channels with non-constant cross-sectional area
or/and non-constant cross-sectional shape. In fact, convergence-divergence
and corrugation (which we mentioned in the previous point) may be
considered as instances or special cases of this point.
\item Flow in curved conduits of the previous geometries (as well as circular
geometry) such as curved pipes of elliptical or rectangular cross
section.
\item Flow in conduits with combinations of the considerations of the previous
points such as the flow in a curved pipe of triangular cross section
with non-constant cross-sectional area.
\end{enumerate}

\subsection{Non-Newtonian Flow in Non-Rigid Structures}

Again, we repeat what was said in the last subsection that is this
issue is not limited to the non-rigid structures of the flow containers
and conduits but to the non-rigid structures plus the attached conditions
(regarding for example the flow, fluid and environment) as well as
the fluid models which are supposed to represent the rheology of the
fluid flowing in these structures.\footnote{In fact, this should apply to all the issues discussed in this section
and hence we will not repeat this note in the following subsections.}

A sample of suggested areas of research in this field is:
\begin{enumerate}
\item Flow in elastic conduits of various geometries, conditions and models
(as discussed in the previous subsection).
\item Flow in viscoelastic conduits of various geometries, conditions and
models (as discussed in the previous subsection).
\item Pulsatile flow in distensible conduits in general (elastic or viscoelastic).
\item Peristaltic flow in distensible conduits in general.
\item Pressure wave propagation through distensible conduits in general.
\end{enumerate}

\subsection{Non-Newtonian Flow in Porous Media}

In fact, this could be included in the previous subsections (mainly
$\S$ \ref{subsecGeometry}) but because of its importance we put
it in a separate subsection. Although there are many investigations
about this subject, there are many gaps and virgin areas of research
in the investigation of non-Newtonian fluid flow in porous media.\footnote{Most of the past and current investigations in the fluid flow through
porous media are either about the flow of Newtonian fluids, or they
use rather simplistic methodologies such as the use of network modeling
with simple geometries and topologies for the porous media elements
(mainly nodes and throats). Moreover, they generally employ simple
flow and ambient conditions and usually ignore rather complex considerations
(such as heterogeneity and anisotropy of porous media or its interaction
with the flowing fluid or its distensibility).}

A sample of suggested areas of research in this field is:
\begin{enumerate}
\item Flow in various ordered and disordered porous media such as 2D and
3D porous media of regular and irregular structure and morphology
(e.g. fractal or lattice-based with and without fixed geometry or/and
size of throats and their connectivity).\footnote{In fact, even entirely irregular (or chaotic) porous media should
be considered in this context.}
\item Flow in distensible porous media of elastic or viscoelastic properties.\footnote{Biological tissues (or some of their types) can be considered as distensible
porous media (noting that biological fluid systems will be discussed
later on; see $\S$ \ref{subsecBiological}).}
\item Flow in fractured porous media.
\item Flow in heterogeneous porous media.
\item Flow in anisotropic porous media.
\item Multi-phase flow in porous media (also see $\S$ \ref{subsecMultiPhase}).
\item Interaction between flowing fluids and porous media (especially distensible
porous media).
\end{enumerate}

\subsection{Non-Newtonian Flow in Micro- and Nano-Scale Systems}

Most of the past and current research in non-Newtonian fluid mechanics
is about macroscopic fluid systems. In fact, this is understandable
since micro- and nano-scale studies (whether in fluid mechanics or
in general) are relatively new. So in short, while the behavior of
non-Newtonian fluids in macroscopic systems is widely studied, the
behavior of such fluids at micro- and nano-scales is generally under-researched.

A sample of suggested areas of research in this field is:
\begin{enumerate}
\item Investigating how non-Newtonian characteristics and attributes (such
as shear-thinning, shear-thickening, viscoelasticity, thixotropy,
yield stress, and so on) change or adapt at reduced scales, especially
under varying structure and ambient conditions (such as confinement,
temperature, and surface effects).
\item Investigating how non-Newtonian behavior can be exploited in new scientific
and technological applications at microscopic and sub-microscopic
scales.\footnote{In fact, fluid flow (and non-Newtonian in particular) at micro- and
nano-scales is important for many traditional and newly-emerging high-tech
technologies. For example, understanding how non-Newtonian fluids
behave in confined environments (such as microfluidic devices or nano-scale
channels) is essential for advancing technologies in drug delivery,
lab-on-a-chip devices, and miniaturized systems.}
\item Investigating the validity of the continuum (or large-scale or bulk)
fluid models (such as power law or Herschel-Bulkley or Carreau) in
micro- and nano-scale systems (noting, for instance, that most non-Newtonian
fluids, such as polymeric liquids, have molecular chain structures
within fluid mediums).
\end{enumerate}

\subsection{Non-Time-Independent Non-Newtonian Flow}

So far, most of the investigations and research in non-Newtonian fluid
mechanics and dynamics are focused on the time-independent fluid flow.
This is understandable due to the relative simplicity of time-independent
non-Newtonian flow and the huge complexities associated with non-time-independent\footnote{We use ``non-time-independent'' instead of ``time-dependent''
to avoid potential suggestion of limitation to thixotropy (which is
usually labeled as time-dependent).} flow (largely classified under the two broad terms of thixotropy
and viscoelasticity). In fact, non-time-independent non-Newtonian
effects and attributes (such as memory effects) are very important
in many natural phenomena and technological applications and hence
they require more attention and investment is research.

A sample of suggested areas of research in this field is:
\begin{enumerate}
\item Investigation of various aspects, attributes and phenomena of viscoelasticity
such as the rate of strain in viscoelastic fluids (and its dependencies
especially on time), hysteresis in stress-strain correlations, storage
modulus, loss modulus, creep, stress relaxation, memory effects in
general, and so on.
\item Investigation of various aspects, attributes and phenomena of thixotropy
such as the rate of strain in thixotropic fluids (and its dependencies
especially on time),\footnote{In fact, time dependencies can be classified into two main types:
intrinsic due to the thixotropic nature and extrinsic due to external
variation of applied stress by the acting agent. This should also
apply to time dependencies of viscoelastic fluids which we discussed
in the previous point.} reversibility in thixotropic cycles, the factors that influence the
thixotropic index, memory effects in general, and so on.
\item Investigation of non-time-independent non-Newtonian effects and attributes
(mainly related to viscoelasticity and thixotropy) at microscopic
and sub-microscopic levels to understand the underlying mechanisms
of these phenomena and their aspects and attributes.
\item Employing and developing more realistic fluid models in the investigation
and modeling of non-time-independent non-Newtonian fluid flows.
\item Investigating various aspects related to non-time-independent non-Newtonian
effects and attributes (such as the influence of memory effects on
heat transfer and flow stability).
\item Investigating how non-time-independent non-Newtonian behavior can
be exploited in new scientific, medical and technological applications.\footnote{Some potential applications that can benefit from the investigation
of non-time-independent non-Newtonian effects and attributes are:
biomedical applications (such as blood flow and related aiding devices),
3D printing, oil extraction and refinement, and industrial manufacturing
in general.}
\item In fact, even some topics of yield stress fluids\footnote{Actually, this may apply to non-Newtonian aspects and attributes other
than yield stress which are commonly and primarily classified as time-independent
behavior.} (such as the flow of these fluids in complex structures where yield
can have complex time-dependent behavior pattern) could be included
(although yield stress is generally considered as a time-independent
effect).
\end{enumerate}

\subsection{Some Types of Time-Independent Non-Newtonian Fluids and Behavior}

In the previous subsection we discussed the research areas of non-time-independent
non-Newtonian fluid mechanics that require more attention and investment.
This may wrongly suggest that the topics of time-independent non-Newtonian
fluid mechanics are properly and sufficiently investigated and researched.
However, this is not true in general since some of the topics of time-independent
non-Newtonian fluid mechanics are not given sufficient attention in
comparison to other (or corresponding) topics of time-independent
non-Newtonian fluid mechanics.

For example, shear-thickening fluids and behavior are not given the
same level of attention as shear-thinning fluids and behavior. Although
this may be justified by the fact that shear-thinning fluids and behavior
are more common, this is not an excuse from a theoretical viewpoint
(and even from some practical viewpoints where shear-thickening fluids
and behavior can potentially have some important uses and applications
that can match or even exceed the importance of uses and applications
of shear-thinning fluids).

Another example is yield stress fluids and behavior. Although there
is a considerable amount of literature about yield stress fluids (and
materials in general) and their behavior and attributes (as well as
uses and applications), there are still many aspects of yield stress
and related phenomena which are not well understood and hence yield
stress and related aspects and phenomena require further attention
and investment in research.\footnote{For example, more research is required in modeling yield stress fluids
and behavior in ordered and disordered porous media so that the yield
point (for instance) of flow in such media can be predicted. Also,
the development (and time dependencies) of flow after yield should
be investigated thoroughly so that they can be predicted and comprehended.}

In short, leveling up in research in these topics (and other similar
topics) of time-independent non-Newtonian fluid mechanics is required
despite the fact that time-independent non-Newtonian fluid mechanics
was (and is still) an area of active and extensive research (at least
since the mid-twentieth century).

\subsection{Non-Newtonian Behavior Under High Strain Rates}

Most of the existing studies of non-Newtonian fluid mechanics are
related to low and medium rates of strain. This is due first to the
relative simplicity of these investigations in comparison to the corresponding
investigations at high strain rates (where many complications arise
at high strain rates), and second to the fact that non-Newtonian fluid
systems at high strain rates are relatively rare in comparison to
those at low and medium strain rates.

Anyway, the behavior of non-Newtonian fluids at high strain rates
(such as during impacts or shock loading) especially with regard to
shear-thickening and memory fluids is an area of research that requires
more attention and investment. This demand is amplified by the fact
that these investigations are essential for understanding impacts,
crash dynamics, flows and ejections at high rates of strain, and some
advanced manufacturing processes (as well as many other natural phenomena
and industrial processes).

A sample of suggested areas of research in this field is:
\begin{enumerate}
\item Developing more accurate and realistic models for these extreme conditions
since most of the existing non-Newtonian fluid models (whether time-independent
or not) are not capable of dealing with the complexities of non-Newtonian
behavior at high rates of strain.
\item Investigating the transitions between different flow regimes to develop
a better understanding of these transitions (and related processes
and phenomena).
\item Investigating how non-Newtonian behavior at high rates of strain can
be exploited in new scientific and technological applications (as
well as in improving our understanding of many natural physical phenomena
and processes).\footnote{For example, a better understanding of non-Newtonian behavior at high
rates of strain can help in industries ranging from automotive to
aerospace, where non-Newtonian fluids are used for damping, shock
absorption, or structural protection (as well as many other processes).
It can also help in understanding many natural physical phenomena
such as lava flow dynamics, eruption styles and magma transport in
volcanoes (as well as some geological and astronomical processes related
to non-Newtonian fluid dynamics).}
\end{enumerate}

\subsection{\label{subsecBiological}Non-Newtonian Fluids in Biological and Biomedical
Systems}

Most of the past and current research in biological and biomedical
fluid mechanical systems (such as blood circulation system and related
devices) is based on Newtonian fluid model and assumptions. Although
this is valid in many instances (either exactly in some cases and
circumstances or approximately in many cases and circumstances) it
is not valid in general especially in some circumstances, systems
and under certain conditions. In fact, the rheology of most biological
fluids (such as blood, mucus, saliva, lymph and synovial fluids) as
well as many fluids used in biomedical applications and devices (such
as gels and oils) is non-Newtonian. While this is well-known, there
is still much to be explored in the context of how these fluids behave
in varying normal and pathological conditions (e.g. diseases, mechanical
stress, varying strain rates, varying temperature and pressure, and
so on).

A sample of suggested areas of research in this field is:
\begin{enumerate}
\item Developing and adopting (and even adapting existing) non-Newtonian
fluid models that can cope with the specialties of biological fluid
systems (noting that most of the existing non-Newtonian fluid models
are developed for industrial and non-biological fluid systems).
\item Investigating how non-Newtonian behavior can be exploited in improving
our understanding of biological phenomena and medical situations.\footnote{An example is investigating how the rheological properties of non-Newtonian
biological fluids change in pathological states (like in the case
of blood during atherosclerosis or mucus during cystic fibrosis) and
how this change could have important implications and impacts on pathological
symptoms, medical diagnostics and treatment design.}
\item Investigating how non-Newtonian behavior can be exploited in biomedical
applications such as developing medical devices and medical procedures.\footnote{An example is investigating how certain non-Newtonian attributes (such
as shear-thinning, yield stress and viscoelasticity) can be exploited
maximally and beneficially in the drug delivery devices and procedures.}
\end{enumerate}

\subsection{\label{subsecMultiPhase}Multi-Phase Non-Newtonian Flow}

Multi-phase non-Newtonian flow\footnote{We should note that ``multi-phase non-Newtonian flow'' should include
the cases where only some phases are non-Newtonian as well as the
cases where all phases are non-Newtonian.} (and actually multi-phase flow in general) is one of the areas of
research that is not investigated properly and sufficiently (due mainly
to its difficulty and complexity)\footnote{In fact, multi-phase flow in general and non-Newtonian in particular
are notorious for their difficulties and complexities. This may be
inferred from the fact that most of the existing studies about multi-phase
flow (including non-Newtonian) are of numerical and computational
nature (e.g. lack of decent and sufficient analytical and experimental
approaches and efforts in this regard) and they are based on many
simplifications and approximations.} and hence it requires further attention and investment. Multi-phase
flow in general and non-Newtonian in particular have important scientific
instances and technological applications such as in biological and
biomedical systems, food manufacturing and processing, and oil recovery
and refinement.

A sample of suggested areas of research in this field is:
\begin{enumerate}
\item Developing novel models and effective strategies (whether analytical
or numerical and computational) for tackling multi-phase flow in general
and non-Newtonian flow in particular.
\item Giving observational and experimental investigations about multi-phase
flow more attention and resources (by encouraging and funding these
investigations).
\item Investigating multi-phase flow in various contexts and situations
such as in biological versus non-biological systems and in porous
media versus conduits of rather simple geometries (such as pipes of
circular and elliptical cross sections, thin slits, and open channels
of rectangular cross section).
\item Investigating multi-phase flow under various conditions and with various
associations such as at extreme ambient conditions (e.g. temperature
and pressure) or with and without flow stimulation (e.g. by mechanical
vibration).
\item Investigating various issues and situations associated with and related
to multi-phase flow such as heat transfer and interfacial interaction
in multi-phase flow.
\item Investigating how non-Newtonian effects (such as shear-thickening,
yield stress and viscoelasticity) affect multi-phase flow (by comparing
Newtonian and non-Newtonian flows in corresponding situations to assess
the impact of the non-Newtonian effects on multi-phase flow).
\item Investigating how non-Newtonian behavior in multi-phase flow can be
employed scientifically and exploited technologically and industrially.
\end{enumerate}

\subsection{Flow Stability and Transition in Non-Newtonian Fluid Dynamics}

There are many virgin (or under-explored) issues and areas of research
related to the stability of flow and the transition between its different
phases and stages. For example, the transition between laminar and
turbulent flow in non-Newtonian fluids is still not fully understood,
particularly when complicated non-Newtonian effects (such as shear-thickening
or viscoelasticity or thixotropy) come into play in this transition.
In fact, these issues and areas of research have many theoretical
and practical benefits and applications. For instance, investigating
flow instabilities and transition mechanisms in non-Newtonian fluids
can shed light on many scientific issues and industrial processes,
such as mixing, pumping, and heat transfer which often rely on and
are affected by flow stability and phase transitions.

\subsection{Rheology of Non-Newtonian Fluids Under External Fields}

There is an emerging interest in how non-Newtonian fluids behave under
the influence of external fields (such as gravitational or electric
or magnetic fields). In fact, this research is still in its infancy.
Investigating how external fields influence the micro-structures,
rheological properties, and flow behavior of non-Newtonian fluids
could have theoretical and practical applications in areas like astronomy,
smart fluids, sensors, and new material design.

\subsection{Other Areas}

There are many other virgin and insufficiently-explored areas of research
in non-Newtonian fluid mechanics which may not be included primarily
under the broad titles of the previous subsections. Some examples
of these areas are:
\begin{enumerate}
\item Non-Newtonian effects in turbulent flow. In fact, turbulent flow in
general (even of Newtonian fluids) is a very challenging and subtle
subject and hence it requires more attention and investment in research.
\item Interaction of Non-Newtonian effects with other fluid dynamics effects
and situations such as slip at fluid solid interface and multi-phase
flow.
\item Technological and industrial applications. There are many technological
and industrial applications of non-Newtonian flow behavior which are
not properly and sufficiently explored and hence they require more
attention and investment in research. An example is non-Newtonian
fluids behavior in additive manufacturing (which is commonly known
as 3D printing) where the use of non-Newtonian fluids in these manufacturing
processes (such as shear-thinning, shear-thickening and viscoelastic
materials) is still an emerging field. The flow behavior during extrusion,
deposition, and solidification in these processes (and similar industrial
processes) is not fully understood. Understanding the interplay between
rheology and printing parameters (such as speed, temperature, and
layer thickness) could lead to the design of more efficient 3D printing
techniques, especially for complex materials (such as hydro-gels and
bio-materials).
\item Non-Newtonian fluids in high energy systems. Non-Newtonian fluids
are used in some high energy technologies and processes (such as geothermal
drilling and hydraulic fracturing), but the fluid dynamics under extreme
conditions (mainly temperature and pressure) and at large-scale is
not fully understood. Developing analytical and computational models
to describe and simulate non-Newtonian fluid behavior in such high
energy systems and under extreme environmental conditions (such as
in deep-sea and deep-well environments) could provide significant
advancements in these high energy technologies. Experimental work
in these areas should also be beneficial to such objectives.
\end{enumerate}

\section{\label{secNovelTools}Introducing Novel Tools and Methods}

In this section we briefly discuss (in broad and general terms) the
novel tools and methods that could and should be introduced in the
research of non-Newtonian fluid mechanics in general (and in the research
of virgin and under-explored areas in particular). In fact, introducing
such tools and methods is not only beneficial to this research but
in many cases it is a necessity and demand.

\subsection{Novel Analytical Methods and Strategies}

It is important to consider developing novel analytical methods and
strategies to deal with the difficult problems and subtle issues of
non-Newtonian fluid mechanics. In fact, this should include the adoption
and use of traditional analytical methods and strategies which are
ignored or not given sufficient attention in the past and current
research of non-Newtonian fluid mechanics. For example, more attention
should be given to the use of variational approaches and optimization
methods in tackling some of the problems and issues of non-Newtonian
fluid mechanics in general and the novel and under-investigated ones
in particular (see for example $\S$ \cite{SochiVariational2013,SochiPresSA22014}).

\subsection{Novel Computational Methods and Strategies}

It is important to consider developing novel computational and numerical
methods and strategies (as well as adopting and using traditional
computational and numerical methods and strategies which are ignored
or not given sufficient attention in the past and current research)
to deal with the difficult problems and subtle issues of non-Newtonian
fluid mechanics in general and the novel and under-investigated ones
in particular. The obvious example in this regard is artificial intelligence
where machine learning techniques (for instance) can be used for more
efficient description and prediction of flow properties and attributes.
In fact, artificial intelligence can be a great aiding tool for dealing
with highly complex situations which are difficult or impossible to
deal with by using traditional methods. Moreover, it can be an aiding
tool even in developing novel analytical methods and strategies (which
we discussed in the previous subsection).

\subsection{Novel Experimental Methods and Strategies}

Again, it is important to consider developing novel experimental methods
and strategies (as well as adopting and using traditional experimental
methods and strategies which are ignored or not given sufficient attention
in the past and current research) to deal with the difficult problems
and subtle issues of non-Newtonian fluid mechanics in general and
the novel and under-investigated ones in particular. In fact, more
investment in experimental work is required in the research of non-Newtonian
fluid mechanics in general (noting that the past and current research
in this field is overwhelmingly theoretical and computational).

\section{\label{secLimitationsOfPast}Limitations of Past and Current Research}

In this section we highlight some of the limitations and shortcomings
of the past and current research in non-Newtonian fluid mechanics.
These limitations and shortcomings should have a special impact on
any potential research in virgin and under-investigated areas in this
field.

\subsection{Limitations of Fluid Models}

There are many limitations in the non-Newtonian fluid models and methodologies
which are developed and used in the past and current research in this
field. Some examples of these limitations are:
\begin{enumerate}
\item Most of the developed and employed non-Newtonian fluid models in the
past and current research in this field are simple and rather unrealistic
(such as power law model). Novel research should address this issue
by developing and using more realistic (and usually more complicated)
non-Newtonian fluid models. This particularly true for the investigation
of processes and phenomena that involve significant non-Newtonian
effects of non-time-independent nature such as viscoelasticity and
thixotropy (as will be outlined in the next point).
\item Most of the developed and employed non-Newtonian fluid models in the
past and current research in this field are about time-independent
flow. Novel research should address this issue by developing and using
non-time-independent flow models.
\item Most approaches in tackling non-Newtonian flow phenomena adopt continuum
models (where the fluids are generally treated as consisting of a
single homogeneous phase). Although these models are good in many
cases and circumstances, they are not realistic in general (for example
in modeling some polymers or some biological fluids in certain biological
systems). Novel research should address this issue by developing and
using non-continuum models to deal with situations where the continuum
models are not appropriate.
\end{enumerate}

\subsection{Limitations of Conditions and Assumptions}

Most of the past and current investigations in non-Newtonian fluid
mechanics are based on simplistic and rather unrealistic conditions
and assumptions with regard to the flow, fluid, conduit and environment
(such as laminarity of flow, incompressibility of fluid, and simplicity
of geometry of flow conduit and ambient conditions like being steady-state
and isothermal). Although such simple conditions and assumptions are
sufficient in many situations (especially those of practical use),
they are not acceptable in general where the situation is too complicated
to be modeled appropriately by such conditions and assumptions. The
following are a few examples of complicated cases and situations that
require more complex and realistic conditions and assumptions:
\begin{enumerate}
\item Non-Newtonian flow under extreme physical conditions (such as high
pressure and temperature).
\item Non-Newtonian flow in dynamical mechanical environment and physical
conditions (e.g. in vibrating vessels and rotating conduits or in
non-adiabatic and non-isothermal conditions).
\item Non-Newtonian flow in conduits with peristaltic movement.
\item Non-Newtonian flow in porous media of complex morphology and topology.
\item Non-Newtonian flow with yield stress.
\item Non-Newtonian flow with non-time-independent non-Newtonian effects.
\item Non-Newtonian flow with total or partial wall slip.
\item Non-Newtonian flow in microscopic and sub-microscopic fluid systems.
\item Cases and circumstances that require consideration of effects and
influences beyond the basic flow situation. For example, more complex
and realistic conditions and assumptions are required in the investigation
of issues and aspects required to develop a proper understanding of
the role and impact of boundary layers, flow instabilities, inertial
effects, and the effects of surface roughness in and on non-Newtonian
fluid dynamics.
\end{enumerate}

\subsection{Limitations of Topics and Phenomena}

There are many limitations of the topics and phenomena in the past
and current research in the field of non-Newtonian fluid mechanics.
A small sample of those topics and phenomena which require further
attention to address these limitations and gaps are:
\begin{enumerate}
\item Non-Newtonian fluid-structure interactions.
\item Non-Newtonian flow in astronomical and astrophysical systems.
\item Non-Newtonian flow in turbulent situations.
\item Non-Newtonian flow in many special settings and circumstances such
as non-Newtonian flow phenomena and effects at branching junctions
or around certain geometric objects (e.g. immobile/vibrating/rotating
cylinder of various cross sectional shapes or sphere or other 3D geometric
objects).
\item Shear banding in non-time-independent non-Newtonian fluid flow.
\end{enumerate}

\subsection{Limitations of Experimental Investigations}

As indicated earlier, there are relatively few experimental studies
in non-Newtonian fluid mechanics in comparison to theoretical and
computational studies. This is due to the practical difficulties and
inconveniences which are usually associated with experimental projects
as well as their extra costs and expenses in comparison to corresponding
theoretical and computational studies.\footnote{Actually, there is another important reason for this bias against
the experimental studies that is experimental work is generally seen
(or ``stereotyped'') as inferior to theoretical and computational
work and hence ``great scientists should not get their hands dirty''
as they should leave dirty tasks to lower-rank scientists or even
technicians. This silly stereotype should be changed for the benefit
and advancement of science.} Anyway, leveling up (or at least reducing the gap) is required in
this regard where more efforts and resources should be allocated to
the experimental (including observational) work in this field. For
example, more funding should be allocated to experimental PhD programs
and post-doctoral research projects.\footnote{In fact, we suspect that there are more experimental studies in this
field (as well as many other fields and branches of science and technology)
within the industry and certain governmental departments and agencies
but theses studies are not published due to their sensitive nature
or commercial competition or lack of interest in publicity or lack
of interest in any benefit beyond the direct practical benefit especially
when such benefit requires unduly extra work (or other similar reasons
and factors).}

\section{\label{secConclusions}Conclusions}

We outline in the following points the main achievements and conclusions
of the present paper:
\begin{enumerate}
\item There are many virgin and insufficiently-investigated areas of non-Newtonian
fluid mechanics. In this investigation we drew the attention of the
research community in the subject of non-Newtonian fluid mechanics
to those research areas and zones of investigation that require more
attention and further investment of efforts and resources in the future
research in this subject.
\item In this investigation we identified the main virgin and under-investigated
areas of research in the field of non-Newtonian fluid mechanics (see
$\S$ \ref{secVirginUnderInvestigated}). We also discussed briefly
the novel tools and methodologies that should be considered and introduced
in the research of these virgin and under-investigated areas (and
actually in non-Newtonian fluid mechanics in general; see $\S$ \ref{secNovelTools}).
Similarly, we discussed briefly the main limitations and shortcomings
of the past and current research in non-Newtonian fluid mechanics
(see $\S$ \ref{secLimitationsOfPast}). All these investigations
and discussions should help in tackling the deficiencies and addressing
the shortcomings of the past and current research in non-Newtonian
fluid mechanics.
\item Our investigation in this paper should help to level up the research
in this field and fill the gaps and vacancies in this subject. It
should also help to avoid rumination in the research of this field
where some researchers repeat what have already been done or add very
little novelty to previous research (or ``reinvent the wheel'').
\item Our investigation should also help PhD students and their advisers
(as well as researchers of higher ranks) in identifying PhD programs
(and research projects) that are worthy of investigation and worthwhile
for investment of efforts and resources for the best outcome and advancement
of science and technology (and knowledge and practical benefits in
general) as well as for the best outcome and advancement of their
career.
\item Our investigation should also motivate and inspire those within the
research community who are looking for breakthroughs and new discoveries
in the field of non-Newtonian fluid mechanics (and even beyond). In
fact, many of the aforementioned areas of research offer a mix of
fundamental challenges (theoretical, computational and experimental)
and practical applications where there is considerable room for new
contributions. Exploring these areas (with traditional and novel tools
and methodologies and with elimination or minimization of the aforementioned
limitations) could lead to groundbreaking research that advances the
understanding and application of non-Newtonian fluids across many
fields of science, medicine and technology.
\end{enumerate}
\phantomsection 
\addcontentsline{toc}{section}{References}\bibliographystyle{unsrt}
\bibliography{Bibl}

\end{document}